\documentstyle[12pt]{article}
\def\ie{{\em i.e.}}
\def\eg{{\em e.g.}}
\def\s{{\,\rm s}}

\def\beq{\begin{equation}}
\def\eeq{\end{equation}}

\catcode`\@=11 

\def\VEV#1{\left\langle #1\right\rangle}

\def\lsim{\mathrel{\mathpalette\@versim<}}
\def\gsim{\mathrel{\mathpalette\@versim>}}
\def\@versim#1#2{\vcenter{\offinterlineskip
    \ialign{$\m@th#1\hfil##\hfil$\crcr#2\crcr\sim\crcr } }}
\def\etal{{\em et al.}}
\def\JL{J. L. Lopez}
\def\DVN{D. V. Nanopoulos}

\def\t1{{\tilde 1}}

\def\GeV{\,{\rm GeV}}
\def\TeV{\,{\rm TeV}}

\def\cm{\,{\rm cm}}

\def\to{\rightarrow}
\def\pb{\,{\rm pb}}
\def\fb{\,{\rm fb}}
\def\ipb{\,{\rm pb}^{-1}}
\def\ifb{\,{\rm fb}^{-1}}
\def \mgev   	{$\GeV$}
\def \chichi 	{{\chi}_{1}^{\pm} {\chi}_{2}^{0} }
\def \chione    {{\chi}_{1}^{\pm}}

\def \gluino 	{\tilde{g}}
\def \squark 	{\tilde{q}}
\def \met	{/\!\!\!\!E_{t}}
\def \sp 	{$\;$}
\def\NPB#1#2#3{Nucl. Phys. B {\bf#1} (19#2) #3}
\def\PLB#1#2#3{Phys. Lett. B {\bf#1} (19#2) #3}

\def\PRD#1#2#3{Phys. Rev. D {\bf#1} (19#2) #3}
\def\PRL#1#2#3{Phys. Rev. Lett. {\bf#1} (19#2) #3}
\def\PRT#1#2#3{Phys. Rep. {\bf#1} (19#2) #3}
\def\MODA#1#2#3{Mod. Phys. Lett. A {\bf#1} (19#2) #3}
\def\IJMP#1#2#3{Int. J. Mod. Phys. A {\bf#1} (19#2) #3}
\def\TAMU#1{Texas A \& M University preprint CTP-TAMU-#1}

\begin{document}
\thispagestyle{empty}
\large
\begin{flushright}
CTP-TAMU-19/94\\
June 1994\\
\end{flushright}
\vspace{.2in}
\begin{center}
\begin{LARGE}
{\bf Supersymmetry at the DiTevatron}\\[.2in]
\end{LARGE}
\vspace{1cm}
\begin{large}
T.~Kamon, J.~L.~Lopez, P.~McIntyre, and J.~T.~White\\
\vspace{1cm}
Department of Physics\\
Texas A\&M University\\
College Station, TX 77832--4242\\
\end{large}
\vspace{1cm}
\end{center}
\begin{abstract}
{\large
We study the signals for supersymmetry at the Tevatron and DiTevatron
($\sqrt{s}=4\TeV$) in various well-motivated supersymmetric models. We consider
the trilepton signature in the decay of pair-produced charginos and
neutralinos, the missing energy signature in gluino and squark production, and
the $b\bar b$ signal in the decay of the lightest supersymmetric Higgs boson
produced in association with a $W$ or $Z$ boson. In each case we perform
signal and background studies, using Monte Carlo and/or real data to
estimate the sensitivity to these signals at the Tevatron and DiTevatron
with the Main Injector, for short- and long-term integrated luminosities of
${\cal L}=10$ and $25\ifb$, and $5\sigma$ statistical significance. We conclude
that one could probe chargino masses as high as
$m_{\chi^\pm_1}\sim180\,(200)\GeV$, gluino masses as high as $m_{\tilde
g}\sim450\,(750)\GeV$, and lightest Higgs boson masses as high as
$m_h\sim110\,(120)\GeV$ at the Tevatron (DiTevatron). A high-luminosity option
at the Tevatron ($10^{33}\cm^{-2}\s^{-1}$) may compensate somewhat for the
higher reach of the DiTevatron, but only in the trilepton and Higgs signals.
However, these gains may be severely compromised once the multiple-interaction
environment of the high-luminosity Tevatron is accounted for.}
\end{abstract}

\newpage
\setcounter{page}{1}
\pagestyle{plain}

\normalsize
\baselineskip=14pt

\section{Introduction}

	In the wake of the demise of the Superconducting Super Collider, the high
energy physics community has been seeking the most cost-effective ways to
extend the reach of present facilities for new science.  Three new developments
during the past year suggest a particular opportunity in this regard.  First,
the recent announcement at Fermilab of evidence for the top quark.  Second,
with the results on electroweak and strong gauge couplings from CERN's LEP
experiments and the new result on the top quark, the models of supersymmetry
have become far more predictive and require a spectrum of new particles in the
mass range of 100-1000 GeV.  Supersymmetry uniquely opens the possibility to
directly connect the Standard Model with an ultimate unification of the
fundamental interactions.  Third, the now-mature magnet technology of the SSC
opens an opportunity to double the energy of the Fermilab Tevatron and access
most of this predicted spectrum \cite{vision}. The DiTevatron would have a
collision energy of 4 TeV and a luminosity of $3\times10^{32}\cm^{-2}\s^{-1}$.
It could be realized by installing a single ring of SSC-type magnets in the
existing tunnel, using the exisiting source and using the Tevatron as a
high-energy injector. This DiTevatron design is summarized in Appendix~A.  It
would require no new tunnel construction, no magnet R\&D, and no new detectors.

Another approach to upgrading the Tevatron, increasing its luminosity to
$>10^{33}\cm^{-2}\s^{-1}$, has also been proposed \cite{design}. Luminosity and
energy trade off up to a point in extending the reach of a collider for new
physics.  The recent evidence for the top quark is an example, however, of the
limit of that trade-off.  The evidence became possible because of a series of
successful upgrades of the Tevatron luminosity which finally brought
sufficient event rate to possibly observe the top; but no luminosity upgrades
would have sufficed to find the massive top quark with half the Tevatron
energy.  The ultimate limit for the mass reach of a hadron collider, resulting
from the distribution functions of the constituent quarks and gluons, is
$\sim25\%$ of its collision energy.  Thus the Sp$\bar{\rm p}$S approached its
limit in the discovery of the $W$ and $Z$, and the Tevatron is approaching its
limit with the evidence for the top quark at 174 GeV.

	The purpose of this paper is to analyze the discovery potential of energy and
luminosity upgrades of the Tevatron. The key question is whether one of these
modest upgrades could provide a major window on new physics during the coming
decade while CERN's Large Hadron Collider (LHC) is being built.

One main motivation for considering an upgraded Tevatron is to study the
physics of supersymmetry. The generic Minimal Supersymmetric Standard Model
(MSSM) is described in terms of a large number of
parameters (at least twenty), which makes experimental tests of such a model
rather impractical. Alternatively, one may consider a theoretical ``framework"
to reduce the number of free parameters, \eg, grand unification, supergravity,
or superstrings. Clearly, the more theoretical assumptions one builds in, the
less parameters the models have, and the more predictive they become.
Even though it is not clear which framework one should consider, once such
a framework is selected, the parameter spaces of these models can be tested
experimentally, either directly through collider processes or indirectly
through rare processes. For concreteness we consider here a four-parameter
``conservative" framework based on supergravity grand unified models with
universal soft-supersymmetry-breaking and radiative electroweak symmetry
breaking. We also study more ``speculative" models (minimal $SU(5)$
supergravity and string-inspired no-scale $SU(5)\times U(1)$ supergravity)
which have much smaller parameter spaces and where one finds an array of
further phenomenological constraints.

Within the context of these models we study three signals for new physics:
(a) the trilepton signature in the decay of pair-produced charginos
($\chi^\pm_1$) and neutralinos ($\chi^0_2$) (section 3.1), (b) the missing
energy signature in gluino ($\tilde g$) and squark ($\tilde q$) production
(section 3.2), and (c) the $b\bar b$ signal in the decay of the lightest
supersymmetric Higgs boson ($h$) produced in association with a $W$ or
$Z$ boson (section 3.3). In each case we perform signal and background
studies, using Monte Carlo and/or real data to estimate the sensitivity to
these signals at the Tevatron and DiTevatron with the Main Injector, for a
short- (long-) term integrated luminosity of ${\cal L}=10\,(25)\ifb$ and
$5\sigma$ statistical significance. These sensitivity results are realistic and
largely model independent. We then obtain the corresponding reaches in
chargino, gluino, and Higgs-boson masses in the models we consider.

Finally we contrast the discovery potential of the Tevatron versus the
DiTevatron in the various luminosity scenarios being considered (section 4). We
conclude that the energy upgrade to the DiTevatron is the most profitable
alternative for the search for supersymmetry at Fermilab.

\section{The models}
\subsection{Conservative framework}
In this case we assume that the models contain the particle content of the
MSSM: the Standard Model particles and their superpartners, plus
two Higgs doublets. Convergence of the precisely-measured Standard Model gauge
couplings (with a suitably normalized hypercharge) then occurs at a scale
$M_U\sim10^{16}\GeV$ \cite{EKN}. From the theoretical point of view, the
unification of the gauge couplings is built into the grand unified model
and the actual experimental test is the predicted value of $\sin^2\theta_W$ in
terms of the strong coupling. Alternatively one can predict the strong coupling
given $\sin^2\theta_W$. All of these tests agree very well with the data.
Supergravity is then invoked as the source of the supersymmetry breaking
scalar and gaugino masses. The simplest assumption is that at the unification
scale supersymmetry is broken in a hidden sector with all scalar masses
degenerate ($m_0$), as are the gaugino masses
($m_{1/2}$), and the trilinear scalar couplings ($A$). The set of mass
parameters is then evolved down to the electroweak scale via the
renormalization group equations, and the whole supersymmetric and Higgs-boson
spectrum is determined (\eg, $m_{\tilde g}\propto m_{1/2}$, $m^2_{\tilde
q}\approx m^2_0+c_{\tilde q}m^2_{1/2}$, and $A$ contributes to the $\tilde
t_L-\tilde t_R$ mass splitting). The final step is to enforce the radiative
breaking of the electroweak symmetry, which allows the determination of the
Higgs mixing parameter $\mu$ (up to a sign).(For recent reviews of this
procedure see Ref.~\cite{reviews}.) The final parameter set also includes the
ratio of the Higgs vacuum expectation values ($\tan\beta>1$), and is
constrained by the present experimental lower bounds on sparticle and
Higgs-boson masses. Incidentally, for users of ISAJET V7.0x \cite{Isajet}, two
of the input parameters are determined in these models, namely $|\mu|$ and the
pseudoscalar Higgs-boson mass ($m_A$). The ISAJET parameter $A_t$ should not
be confused with the parameter $A$, which is the value $A_t$ takes at the
unification scale. In what follows we take $A=0$, which nonetheless implies
a non-zero value for $A_t$.

The top-quark mass is an essential input in the calculations, although small
variations do not affect the results significantly; we take $m_t^{\rm
pole}=174\GeV$ \cite{CDF} in discussions of this model. Our exploration of the
four-dimensional parameter space is necessarily a limited one:
$\tan\beta=2,10$; $\xi_0\equiv m_0/m_{1/2}=0,1,2,5$; $A=0$, and a variable
chargino (\ie, $m_{1/2}$) mass. (As shown in Eq.~(\ref{sq-gl}) below,
$\xi_0=0,1,2,5\leftrightarrow m_{\tilde q}\sim(0.8,0.9,1,2)m_{\tilde g}\,$.)

\subsection{More speculative models}
We would also like to consider models with further theoretical and
phenomenological constraints, which reduce the size of the allowed parameter
space. These models are particular cases of the conservative models described
above.

In the minimal $SU(5)$ supergravity model \cite{Dickreview}, specification of
the GUT gauge group entails two new phenomenological constraints: (i) proton
decay via $p\to\bar\nu K^+$ \cite{LNP+LNPZ+AN}, which entails small values of
$\tan\beta\ (\lsim10)$, relatively light charginos and gluinos, and relatively
heavy squarks and sleptons (\ie, $\xi_0\gsim3$); (ii) the relic density of the
lightest neutralino should not be too large \cite{cosmo}, which in conjunction
with the proton decay constraint results in $m_{\chi^0_1}\sim{1\over2}m_h$ or
$\sim{1\over2}m_Z$. $SU(5)$ symmetry also implies unification of the bottom and
tau Yukawa couplings, which favors $\tan\beta\sim1$ \cite{yuks} (or
$\tan\beta\gg1$). We do not impose this condition here as it is sensitive to
relatively small perturbations that could arise from Planck-scale physics. We
choose here $m^{\rm pole}_t\approx168\GeV$. The combined constraints of (i) and
(ii) imply
\begin{equation}
m_{\chi^\pm_1}\lsim120\GeV,\quad m_{\tilde g}\lsim400\GeV,\quad
m_h\lsim120\GeV.
\label{minranges}
\end{equation}

In the string-inspired $SU(5)\times U(1)$ supergravity model \cite{Faessler}
there are intermediate scale particles (at $\sim10^6\GeV$ and
$\sim10^{12}\GeV$) which delay unification until the string scale ($M_{\rm
string}\sim10^{18}\GeV$). We also consider the no-scale supergravity
\cite{LN,LNZI} universal soft-supersymmetry-breaking scenario with $m_0=A=0$
(\ie, $\xi_0=0$). Thus, this is a two-parameter model ($\tan\beta$ and
$m_{1/2}\leftrightarrow m_{\chi^\pm_1}\leftrightarrow m_{\tilde g}$), which has
been studied in Refs.~\cite{LNZI,Easpects}, including additional indirect
experimental constraints (\eg, from $b\to s\gamma$ and $(g-2)_\mu$ processes).
We note that in $SU(5)\times U(1)$ supergravity the proton decay mode
$p\to\bar\nu K^+$ is automatically small, the cosmological relic
density is always below cosmological limits, and no Yukawa unification
condition is required by the $SU(5)\times U(1)$ gauge symmetry.

\section{Signals for supersymmetry}
We now discuss three typical signals for supersymmetry in the models discussed
above at the upgraded Tevatron, namely chargino-neutralino production and
decay via the trilepton channel, the missing energy signature in squark and
gluino production, and associated production of the lightest Higgs boson. These
are not the only possible signals, but we believe they are the most important
ones. The analysis is facilitated considerably because of the definiteness of
the parameters to be explored and the relationships among the various sparticle
masses. The latter will be discussed in the following subsections as needed.

\subsection{Charginos and neutralinos}
In the models we consider, a simple relation among the lighter neutralino and
chargino masses holds to varying degrees of accuracy, namely
$m_{\chi^0_2}\approx m_{\chi^\pm_1}\approx2 m_{\chi^0_1}$ \cite{ANc,LNZI}. The
process of interest: $p\bar p\to \chi^0_2\chi^\pm_1X$, where both
neutralino and chargino decay leptonically ($\chi^0_2\to\chi^0_1 \ell^+\ell^-$,
$\chi^\pm_1\to \chi^0_1 \ell^\pm\nu_{\ell}$, with $\ell=e,\mu$) was first
treated for on-shell $W$'s in Ref.~\cite{EHNS+others}. The production cross
section for off-shell $s$-channel $W$-exchange and $t$-channel
squark-exchange (a small contribution for heavy squarks), was first
studied at the Tevatron in Refs.~\cite{trileptons,BaerTata}, and has also been
explored in $SU(5)\times U(1)$ supergravity in Ref.~\cite{LNWZ}.
In figures \ref{DiTev.SSM2} ($\tan\beta=2$) and
\ref{DiTev.SSM10} ($\tan\beta=10$) we give the cross sections into trileptons
(summed over all four channels: $eee,ee\mu,e\mu\mu,\mu\mu\mu$) at the Tevatron
and DiTevatron, versus the chargino mass for the conservative models discussed
above. In figures \ref{DiTev.min} and \ref{DiTev.nsc} we show the analogous
results for the minimal $SU(5)$ and no-scale $SU(5)\times U(1)$ supergravity
models. Before we can assess the discovery potential of these models via
the trilepton signature, we have to discuss the experimental reach in various
luminosity scenarios.

	The CDF and D0 collaborations have collected about $20\ipb$ of data in the
1992--93 run. The data analysis on supersymmetry searches using trilepton
events sets an upper limit of about $2\pb$ for
$\sigma(\chi^\pm_1\chi^0_2)\times
B$ into all four trilepton modes ($eee,ee\mu,e\mu\mu,\mu\mu\mu$)
\cite{White,Kato}. The major backgrounds are $t \bar{t}$, $ZW$, $ZZ$,
$Z+X$ and ${\rm DY}+X$, where $X$ could be a real lepton or a fake lepton.
Hereafter, the combined lepton contribution is called ``fake". The acceptance
for the events is calculated using ISAJET(V7.06) + QFL (a CDF detector
simulation program). This includes the detector smearing effect,
inefficiency due to uninstrumented regions of the detector, and lepton
identification efficiency. We also assume that (a)~the beam luminous region is
a Gaussian distribution with a sigma of 30~cm and that (b)~the coverage for
leptons in the future upgraded detector will be the same as in the current
detector. The $P_t$ ($E_t$) cut for the leading muon (electron) in a trilepton
event is required to be $10\GeV$, and the minimum $P_t$ ($E_t$) cut is $4\GeV$
($5\GeV$) for muons (electrons). Additional cuts for
the event selection \cite{Kato} are required:
\begin{itemize}
\item  $| Z_{\rm vertex} | <$ 60~cm,
\item  $\Delta R(\ell \ell) > 0.4$ (for any two leptons),
\item  $Iso(R=0.4) < 2\GeV$ (energy isolation around each lepton),
\item  Unlike-sign requirement ($e^+e^-$ or $\mu^+\mu^-$),
\item  Resonance removal: $Z$ (75--105 GeV),
		$\Upsilon$ (9--11 GeV), and $J/\psi$ (2.9--3.3 GeV),
\item  $\Delta\phi(\ell_1,\ell_2) < 170^\circ$ (for the two leading $P_t$
leptons).
\end{itemize}

The isolation energy cut ($2\GeV$) is determined by looking at the fluctuation
of the energy flow within $R = 0.4$ around the lepton in the data.
The $\Delta\phi$ (azimuthal opening angle) cut on the two leading leptons
is found to be effective in reducing the DY events by a factor of 2,
whereas $\sim10\%$ of the signal events are rejected. Trigger efficiency for
the trilepton event trigger is taken to be 90\% \cite{Kato}. We also assume the
fake lepton rate in $Z$ and DY to be $10^{-4}$. The current CDF and D0 fake
lepton rates are somewhat worse; they should be improved in the future upgraded
detectors.

	The background is estimated as follows:
\begin{eqnarray}
 N_{{\rm BG}}  &=& \sigma \times B_{2\ell} \times {\epsilon}({\rm MC})_{2\ell}
		\times {\rm \epsilon({\rm trig})} \times f
		\times {\cal L},\qquad {\rm for\ {\em Z}\ and\ DY}\\
	      &=& \sigma \times B_{3\ell} \times {\epsilon}({\rm MC})_{3\ell}
		\times {\rm \epsilon({\rm trig})} \times {\cal L},\qquad
				{\rm for\ {\em ZW}\ and\ {\em ZZ}} \\
	      &=& \sigma \times B_{2\ell} \times {\epsilon}({\rm MC})_{3\ell}
		\times {\rm \epsilon({\rm trig})} \times {\cal L},\qquad
				{\rm for\ {\em t \bar{t}} }
\end{eqnarray}
where $\sigma\times B$ is the cross section to produce
dilepton or trilepton final states.
The production cross sections are obtained from
ISAJET + CTEQ2L structure functions
with K-factors of 1.0 for $t \bar{t}$,\footnote{
The production cross sections for $t \bar{t}$ at $\sqrt{s}=$ 2 and 4 TeV
are $7\,(15)\pb$ and $42\,(78)\pb$ for $m_t= 170\,(150)\GeV$ \cite{Isajet}.
The branching ratio for the dilepton mode
($t \bar{t} \rightarrow ee+X, \; e \mu+X, \; \mu \mu+X$) is 4.4\%.}
1.3 for $Z$ and DY, and 1.4 for $ZZ$ and $ZW$.
With these K-factors the cross sections at $\sqrt{s} = 1.8\TeV$ are consistent
with the current CDF data for $Z$ and DY \cite{T3,T4} and theoretical
calculations for $t \bar{t}$ \cite{T9}, and $ZZ$/$ZW$ \cite{T5,T6}.
Also, $\epsilon({\rm MC})_{3\ell}$
($\epsilon({\rm MC})_{2\ell}$) is an acceptance for
trilepton (dilepton) events with the above cuts and
is determined by a Monte Carlo
simulation (ISAJET + QFL), $\epsilon({\rm trig})$ is an expected trigger
efficiency (90\%), $f$ is the fake rate ($f=10^{-4}$), and  ${\cal L}$ is the
integrated luminosity.

The $t \bar{t}$ dilepton events could be an important background to the
trilepton signal \cite{trileptons}. The additional lepton might come from the
$b$-quark leptonic decay, which cannot be rejected with an isolation cut at
certain rate. The fraction of three isolated leptons (with $Iso < 2$ GeV)
in top dilepton events is determined to be $0.63\,(0.54) \times 10^{-3}$ at
2\,(4) TeV. Therefore we expect
$(7\times10^3\fb)(4.4\%)(0.63\times10^{-3})(1\ifb)=0.19$ events with
1~fb$^{-1}$
at $\sqrt{s}$ = 2 TeV, and 1.0 events at 4 TeV. At 4 TeV we can reduce this
background by a factor of 17 by requiring no jets with $E_{t} \geq$ 25 GeV in
$|\eta({\rm jet})| < 2.4$. This cut keeps 70\% of the signal, \ie, a drop
of 30\% in the signal significance. Since, as we show below, the resulting
$t\bar t$ background is only about 25\% of the other backgrounds, it is better
to not impose the jet cut. This entails a degradation of the significance by
only 12\%.

Another possible source of background comes from three-jet events faking a
trilepton signal. Based on the CDF measurements of jet fragmentation \cite{T7},
the probability of one charged track carrying more than 80\% of its jet energy
is less than $10^{-4}$. This can also be used for neutral particles (with a
factor of 1/2). A rough estimate of the probability that such a particle fakes
an $e^\pm$ or $\mu^\pm$ in a magnetic detector (\eg, CDF) is about 1--5\%
(depending on the tightness of the selection criteria). Therefore, we take as a
conservative estimate for the fake rate of a jet being misidentified as $e^\pm$
or $\mu^\pm$ the value $10^{-5}$. Since the three-jet cross section is about 1
mb, the contribution to ``trilepton" events is
$1\,{\rm mb}\times(10^{-5})^3=10^{-3}\fb$.

\begin{table}[t]
\caption{Number of background events expected in trilepton searches in
$p\bar p$ collisions at center-of-mass energy of 2 and 4 TeV for ${\cal
L}=1\,{\rm fb}^{-1}$. $X$ represents a fake lepton.}
\label{Table1}
\begin{center}
\begin{tabular}{|l|l|l|}\hline
Process&$N_{{\rm BG}}(2\TeV)$&$N_{{\rm
BG}}(4\TeV)$\\ \hline
$ZW$&$0.21$&$0.49$\\
$ZZ$&$0.04$&$0.08$\\
$(Z\to \ell \ell)+X$&$0.13$&$0.24$\\
$(Z\to\tau\tau\to \ell \ell)+X$&$0.10$&$0.18$\\
$({\rm DY}\to \ell \ell)+X$&$1.95$&$3.08$\\
$({\rm DY}\to\tau\tau\to \ell \ell)+X$&$0.01$&$0.02$\\
$t \bar{t}$	& 0.19	& 1.00 \\ \hline
Total&2.63 events& 5.09 events\\ \hline
\end{tabular}
\end{center}
\hrule
\end{table}

	The estimated number of background events at
$\sqrt{s} = 2,4\TeV$ with ${\cal L}=1 \ifb$ are given in
Table~\ref{Table1}.\footnote{Here we use $m_t=170\GeV$. Using $m_t=150\GeV$
instead increases the total number of background events by $8\%\,(17\%)$ at
$\sqrt{s}=2\,(4)\TeV$.}
{}From Table~\ref{Table1} we calculate the sensitivity
(minimal observable signal cross section) for trilepton searches for a
$5\sigma$ statistical significance:
\begin{equation}
{\rm Sensitivity}\, [\fb]= {5\times \sqrt{N_{{\rm BG}}}\over
{\cal L}\times {\rm \epsilon({\rm MC})\times \epsilon({\rm trig})}}\ .
\label{sensitivity}
\end{equation}

The value of $\epsilon({\rm MC})$ depends on the chargino mass and on whether
the leptonic decay is dominantly a two-body or a three-body process. For
three-body decays, a typical value is $\epsilon({\rm MC})\times
\epsilon({\rm trig})=0.12\,(0.09)$ at 2 (4) TeV,
which is valid for $m_{\chi^\pm_1}\approx120\GeV$
and does not change much for $m_{\chi^\pm_1}\gsim120\GeV$. We
have found that the acceptance for two-body decays for
$m_{\chi^\pm_1}\gsim120\GeV$ is almost the same as that for the three-body
decays. The reason is that our trigger lepton $P_t$ threshold (10 GeV) is low
enough to detect the highest $P_t$ lepton in two- or three-body decays, and our
minimum lepton $P_t$ cuts (5 GeV for $e$ and 4 GeV for $\mu$) are low enough
for unbiased detection of the other two leptons. Since the $\chi^\pm_1$ and
$\chi^0_2$ are relatively heavy, most leptons tend to be detected in the
central region. Therefore, the acceptance for trilepton events is mainly
limited by the detector coverage (fiducial region in $\eta$ and $\phi$).
In what follows we use the typical value ${\rm \epsilon(MC)\times
\epsilon(trig)}=0.12\,(0.09)$ at 2~(4)~TeV.

The above discussion assumes that the leptons are hard enough to be
detectable, but this may not always be the case when two-body decays dominate.
Kinematically speaking, in the parent rest-frame the $P_t$ cuts entail a
minimum detectable daugther lepton energy ($E^{\rm min}_{\ell}$),
and thus a minimum
mass difference ($\Delta m$) between the chargino (neutralino) and the
sneutrino (selectron) when the two-body decay $\chi^\pm_1\to\tilde\nu e^\pm$
($\chi^0_2\to\tilde e_R e$) is allowed and dominates the decay amplitude
(\ie, $\Delta m\gsim E^{\rm min}_{\ell}$ for parent masses in the range of
interest). In the Lab frame these simple relations are smeared and a simulation
is required to obtain the $P_t$ distribution as a function of $\Delta m$.
As an example, we studied a case (for $\xi_0=0$ and $\tan\beta=2$) with
$m_{\chi^\pm}=112.6\GeV$ and $m_{\tilde \nu}=111.2\GeV$ (\ie, $\Delta
m=1.4\GeV$). In this case the efficiency for trilepton events was found to be
$1\over4$ of the $12\%$ quoted above. This result is encouraging since such
small values of $\Delta m$ occur only rarely in the models we have studied, and
only for $\xi_0=0$. (However, it is precisely for this value of $\xi_0$ that
one gets the largest trilepton rates (see
Figs.~\ref{DiTev.SSM2},\ref{DiTev.SSM10}).)

\begin{table}[t]
\caption{Sensitivity at $5\sigma$ significance for trilepton searches into
$eee,ee\mu,e\mu\mu,\mu\mu\mu$ at the Tevatron and DiTevatron, for
expected short- and long-term integrated luminosities. We assume 0.12~(0.09)
efficiency for trilepton detection at 2~(4)~TeV. The sensitivity at the
Tevatron at the end of Run IB (1994--95, ${\cal L}\sim0.1\ifb$) is expected to
be $\sim400\fb$.}
\label{Table2}
\begin{center}
\begin{tabular}{|c|c|c|c|c|}\hline
$\sqrt{s}$&${\cal L}(\ifb)$&$N_{{\rm BG}}$&$N_{{\rm S}}$&Sensitivity(fb)\\
\hline
2 TeV
&10&26&25&21\\
&25&66&41&14\\ \hline
4 TeV
&10&51&36&40\\
&25&127&56&25\\ \hline
\end{tabular}
\end{center}
\hrule
\end{table}

In Table~\ref{Table2} we summarize the sensitivities for two integrated
luminosity scenarios with and without the energy upgrade. We assume an
instantaneous luminosity of $3\times10^{32}\cm^{-2}\s^{-1}$, which could be
achieved with the Main Injector (see Appendix~A) and is the expected upper
limit allowed by the present CDF and D0 detectors (including their planned
Main-Injector-era upgrades). With a 50\% duty cycle, one would accumulate
$10\ifb$ in two years (short term) and $25\ifb$ in five years (long term). We
do not consider the proposed luminosity upgrade (so-called $\rm T^*$ Tevatron)
to $10^{33}\cm^{-2}\s^{-1}$ since that would require basically new expensive
detectors and a time-line which is beyond the planned start-up time of the
LHC.

\begin{table}[t]
\caption{Reach for chargino masses at $5\sigma$ significance in $p\bar p$
collisions at the Tevatron and DiTevatron for ${\cal L}=10\,(25)\ifb$ in the
conservative models which we consider. All masses in GeV. An asterisk indicates
that the whole allowed range of chargino masses is covered.}
\label{Table3}
\begin{center}
\begin{tabular}{|c|c||c|c||c|c|}\hline
\multicolumn{2}{|c||}{Parameters}&\multicolumn{2}{c||}{Tevatron}&\multicolumn{2}{c|}{DiTevatron}\\ \hline
$\tan\beta$&$\xi_0$&$\mu>0$&$\mu<0$&$\mu>0$&$\mu<0$\\ \hline
&$0$&$125^*\,(125^*)$&$170\,(180)$&$125^*\,(125^*)$&$190^*\,(190^*)$\\
$2$&$1$&$155\,(155)$&$160\,(170)$&$155\,(155)$&$175\,(185)$\\
&$2$&$140\,(150)$&$135\,(145)$&$150\,(155)$&$145\,(160)$\\
&$5$&$115\,(125)$&$115\,(125)$&$120\,(135)$&$115\,(130)$\\ \hline
&$0$&$100\,(110^*)$&$115^*\,(115^*)$&$100\,(110^*)$&$115^*\,(115^*)$\\
$10$&$1$&$125\,(135)$&$135\,(145)$&$135\,(145)$&$140\,(155)$\\
&$2$&$95\,(105)$&$105\,(115)$&$95\,(110)$&$110\,(125)$\\
&$5$&$95\,(105)$&$105\,(115)$&$95\,(110)$&$110\,(125)$\\ \hline
\end{tabular}
\end{center}
\hrule
\end{table}

Taking the sensitivities given in Table~\ref{Table2}, we can determine the
approximate reach in chargino masses for each of the (conservative) models in
the various scenarios by examining figures~\ref{DiTev.SSM2} and
\ref{DiTev.SSM10}. These reaches are given in Table~\ref{Table3}. Note the
many instances where the whole range of chargino masses is accessible
(\ie, the asterisks in Table~\ref{Table3}). In the case of the minimal $SU(5)$
supergravity model, because of the model contraints in Eq.~(\ref{minranges}),
basically the whole range of chargino masses should be accessible at the
Tevatron and DiTevatron with ${\cal L}=25\ifb$, as figure~\ref{DiTev.min}
shows. Note that the slope of the trilepton rate plots is comparable to
the $\xi_0=5$ curves in Fig.~\ref{DiTev.SSM2}, since in this model
$4\lsim\xi_0\lsim10$. This also implies that a limiting point is reached
where increasing $\xi_0$ further does not change the slope of the plots
anymore.

For the no-scale $SU(5)\times U(1)$ supergravity model (see
Fig.~\ref{DiTev.nsc}), aymptotically the trilepton rates are comparable with
the $\xi_0=0$ cases in Fig.~\ref{DiTev.SSM2}, since in this model $m_0=0$.
There are however some differences, especially for lighter values of the
chargino mass. These are due to the somewhat different relationships among the
sparticle masses compared to the conservative models (for $\xi_0=0$), because
of the different unification scales. The reach in this case would be (for
$\mu<0$) $m_{\chi^\pm_1}\approx190\,(200)\GeV$ at the Tevatron and
$220\,(240)\GeV$ at the DiTevatron, for ${\cal L}=10\,(25)\ifb$.

\subsection{Gluino and squarks}
The relationship between gluino and squark masses depends on the parameter
$\xi_0$. In the conservative models discussed above, the average squark mass
is approximately given by (for $\alpha_3=0.120$)
\begin{eqnarray}
m_{\tilde q}&\approx& m_{\tilde g}\left({\sqrt{6+\xi^2_0}\over2.9}\right)
\label{sq-gl}\\
&\approx&0.84\, m_{\tilde g}\qquad {\rm for}\ \xi_0=0\nonumber\\
&\approx&0.91\, m_{\tilde g}\qquad {\rm for}\ \xi_0=1\nonumber\\
&\approx&1.09\, m_{\tilde g}\qquad {\rm for}\ \xi_0=2\nonumber\\
&\approx&1.91\,m_{\tilde g}\qquad {\rm for}\ \xi_0=5\nonumber
\end{eqnarray}

In order to determine the reach of the Tevatron and DiTevatron for squarks and
gluinos, we studied the two extreme cases: (i) gluino pair production with
significantly heavier squarks, \ie, the limit $\xi_0\gg1$ in Eq.~(\ref{sq-gl})
(also expected in the minimal $SU(5)$ supergravity model), and (ii) all gluino
squark production channels such that $m_{\tilde q}\approx m_{\tilde g}-10\GeV$,
as expected for $\xi_0\lsim1$ (also expected in the no-scale $SU(5)\times U(1)$
model).
For case (i), events were generated using ISAJET (V7.06) for different gluino
masses with $m_{\tilde q}= m_{\tilde b}= m_{\tilde t}  = 1 \TeV$.
For case (ii), all squarks were given masses 10 GeV less than the gluino mass
while sleptons and sneutrinos were left at 1 TeV. In both cases,
the following additional parameters were chosen: $\tan{\beta} = 2$, $\mu =
-500\GeV$, $A_t = -100\GeV$, and $m_A = 500\GeV$.

The events were processed through a toy calorimeter using energy smearing with
D0 resolution and jet finding algorithm with a cone with $R =
0.7$. A variety of cuts based on missing $P_t$ and the number of jets above
a given threshold were studied to achieve a reasonable signal-to-background
ratio. Other more refined cuts (such as the direction of the missing $P_t$
vector relative to the leading jets) were not considered. The following
selection criteria were then applied:
\begin{itemize}
\item Tevatron: missing $P_t > 100\,{\rm GeV}$, 3
jets with $P_t > 40\,{\rm GeV}$
\item DiTevatron: missing $P_t > 150\,{\rm GeV}$, 4
jets with $P_t > 40\,{\rm GeV}$
\end{itemize}

For the background estimate, only the dominant $Z \to \nu \nu$ channel was
generated (using ISAJET).\footnote{We have also studied the possible
$t\bar t\to \ell\nu+n-{\rm jets}$ background, where the lepton from $W$ decay
is lost. For $m_t=170\GeV$ we found the $t\bar t$ background to be no larger
than the $Z \to \nu \nu$ background. For completeness, a detector-dependent
discussion is given in
Appendix~B.} For selections with large missing $P_t$, and with the
upgraded detectors, it was assumed that contributions from leptonic $W$ decays
with missed electrons and muons will be no larger than the $Z \to \nu \nu$
background, and that QCD backgrounds are negligible. To account for these and
any other backgrounds, and for any inefficiencies from
additional cuts, the $Z \to \nu \nu$ background was multiplied by a factor of
five. This should provide a conservative estimate of the reach based on
current CDF and D0 experience.

An estimate of the reach with ${\cal L}=10\ifb$ for these two cases is shown in
Fig.~\ref{MonoTev.gluino} for the Tevatron and in Fig.~\ref{DiTev.gluino} for
the DiTevatron, in terms of the statistical significance $N_{\rm
S}/\sqrt{N_{\rm BG}}$. For other values of ${\cal L}$, simply multiply the
significance by a factor of $({\cal L}/10\ifb)^{1/2}$. The reaches that could
be achieved with ${\cal L}=10\,(25)\ifb$ are summarized in Table~\ref{Table5}.

\begin{table}[t]
\caption{Reach for gluino masses at $5\sigma$ significance in $p\bar p$
collisions at the Tevatron and DiTevatron for ${\cal L}=10$ and $25\ifb$ in the
two extreme scenarios: (i) $m_{\tilde q}\gg m_{\tilde g}$ and (ii)
$m_{\tilde q}=m_{\tilde g}-10\GeV$. The ranges indicate the allowed uncertainty
in the background: $(Z\to\nu\bar\nu$) up to $5\times(Z\to\nu\bar\nu)$. All
masses in GeV.}
\label{Table5}
\begin{center}
\begin{tabular}{|c||c|c||c|c|}\hline
&\multicolumn{2}{c||}{Tevatron}&\multicolumn{2}{c|}{DiTevatron}\\ \hline
${\cal L}$&$m_{\tilde q}\gg m_{\tilde g}$&$m_{\tilde q}=m_{\tilde g}-10$
&$m_{\tilde q}\gg m_{\tilde g}$&$m_{\tilde q}=m_{\tilde g}-10$\\ \hline
$10\ifb$&330--370&400--430&540--590&670--720\\
$25\ifb$&360--400&410--440&560--610&690--750\\ \hline
\end{tabular}
\end{center}
\hrule
\end{table}

Note that in the minimal $SU(5)$ supergravity model (where $m_{\tilde q}\gg
m_{\tilde g}$), the whole range of allowed gluino masses (see
Eq.~(\ref{minranges})) could be explored at the DiTevatron. For the case
consistent with $\xi_0\lsim1$ (\ie, $m_{\tilde q}\approx m_{\tilde
g}$), the DiTevatron could explore ultimately roughly 75\% of the parameter
space.

\subsection{Lightest Higgs boson}
For the models we consider in this paper the lightest Higgs
boson is very much Standard-Model--like, and thus the LEP limit
on the Standard Model Higgs-boson mass ($m_H\gsim65\GeV$) applies as well here.
The mass of the lightest supersymmetric Higgs boson ($h$) is bounded above by
an $m_t$-dependent limit: $m_h\lsim120\,(130)\GeV$ for $m_t=150\,(170)\GeV$.
Therefore, signatures for the difficult intermediate-mass Higgs boson need
to be explored. We consider the associated production mechanism $p\bar p\to
W^*,Z^*\to Wh,Zh$ \cite{GNY,others}, which has been recently revisited in
Ref.~\cite{SMW}. The decays of the Higgs boson in this mass range are
dominantly to $b\bar b$ final states, except when the supersymmetric
$h\to\chi^0_1\chi^0_1$ mode is kinematically allowed (in a small region of
parameter space).

	Higgs searches will be in the mainstream of any future collider program. At
the moment only the planned LHC supercollider at CERN could
possibly explore the largest range of Higgs parameter space. However, this will
require ultimate luminosity ($>10^{34}\cm^{-2}\s^{-1}$) and any detector
will suffer from numerous multiple interactions. This makes background
(physics, fake, and maybe a combination of both) analyses difficult. In light
of the proposed Tevatron upgrades \cite{vision,design}, two analyses have
appeared dealing with associated Higgs production and detection
\cite{GH,SMWII}. Moreover,
it has been pointed out that double $b$-tagging reduces the $W+jj/Z+jj$
background substantially, but $W+bb/Z+bb$ and $W+jj/Z+jj$ remain the main
background sources \cite{GH,SMWII}. Since the lightest Higgs boson in the
models of interest here looks very much like the Standard Model Higgs boson,
in what follows we concentrate on the latter.

We have studied the event topology of $H\to b\bar b$ decay using the PYTHIA
Monte Carlo program \cite{pythia} to see if any further useful cuts (beyond
those imposed in Refs.~\cite{GH,SMWII}) may exist that enhance the
signal-to-background ratio. We found that a cut in $\cos(\theta^*)$, where
$\theta^*$ is
the polar angle with respect to the $2b$ (or $2j$) direction in the $b\bar b$
(or $jj$) center-of-mass system, reduces the QCD background while keeping a
large fraction of the signal. A plot of this distribution is shown in
Fig.~\ref{costheta}.  In our study we considered $Z+H\to ee + b\bar b$ and
$Z(\to ee)+b\bar b$ with a smearing of electron and jet energies:
$\sigma/E = 15\%/\sqrt{E}\oplus 1\%$ for electrons and $\sigma/E=
80\%/\sqrt{E}\oplus 5\%$ for jets.
After the smearing the following kinematical and geometical cuts were imposed:
\begin{itemize}
\item	For $b$:     $|\eta(b)| < 2$,     $P_t(b) > 15\GeV$
\item	For $\ell$:  $|\eta(\ell)| < 2$,  $P_t(\ell) > 20\GeV$
\item	Topology cuts:	$\Delta R(b \bar b) > 0.7$, $\Delta R(b \ell) > 0.7$
\end{itemize}
After the event selection, the $\cos(\theta^*)$ cut is imposed
\begin{equation}
		\cos(\theta^*) < 0.7
\end{equation}
in the $b\bar b$ center-of-mass system. This cut accepts 75\% of the signal
(mass-independent) and 35\%-46\% (depending on the Higgs-boson mass) of the
background. Table~\ref{Table4} summarizes the ratio of event acceptance,
\begin{equation}
R={\# {\rm events\ with\ the\ \cos(\theta^*)\ cut}\over
\# {\rm events\ without\ the\ \cos(\theta^*)\ cut}}\ ,
\end{equation}
for background and signal. We also define the improvement factor of the
significance as
\begin{equation}
I = \VEV{ R_{{\rm S}} } / \sqrt{ R_{{\rm BG}} },
\end{equation}
where $\VEV{ R_{{\rm S}} }$ is the average value of $R_{{\rm S}}$. These
numbers are also listed in Table~\ref{Table4}.

\begin{table}[t]
\caption{Improvement factor of $Z+(H\to b\bar b)$ signal over $Z+b\bar b$
background in the presence of the $\cos(\theta^*)$ cut discussed in the text.
Results are for $p\bar p$ collisions at center-of-mass energy of 4 TeV. For
the signal, we get an average $\langle R_{\rm S}\rangle=0.75$. The improvement
factor is defined as $I=\VEV{R_{{\rm S}}}/[R_{{\rm BG}}]^{1/2}$.}
\label{Table4}
\begin{center}
\begin{tabular}{|c|c|c|c|}\hline
$M_{b\bar b}(\GeV)$&$R_{{\rm BG}}$&$R_{{\rm S}}$&$I$\\ \hline
$100\pm20$&0.46&0.724&1.11\\
$110\pm20$&0.40&0.750&1.19\\
$120\pm20$&0.36&0.749&1.25\\
$140\pm20$&0.35&0.744&1.27\\ \hline
\end{tabular}
\end{center}
\hrule
\end{table}

We see that the significance of the signal is improved by 10\%--30\% for
Higgs-boson masses in the range $(100-140)\GeV$, when adding the
$\cos(\theta^*)$ cut. This improvement factor can also be used for $W+H$ over
$W+jj/W+b\bar b$. If we can assume a similar improvement factor for all other
backgrounds ($t\bar t, Wg\to t\bar b, q\bar q\to t \bar b$,
$WZ$~\footnote{For $M_{b\bar b}\gsim110\GeV$ this background is smaller than
all the others (see Table 2 in Ref.~\cite{SMWII} or Fig. 2 in Ref.~\cite{GH}).
On the other hand, for $M_{b\bar b}\approx90\pm20\GeV$, this background is the
dominant one and the $\cos(\theta^*)$ cut is not as effective (\ie, we find
$R_{{\rm BG}}=0.65$).})
then we can push up somewhat the Higgs-boson mass reaches
at the DiTevatron. We start from Table~1 of Ref.~\cite{GH}, where the signal,
background, and statistical significance are given for $m_H=(60-130)\GeV$
at the DiTevatron with ${\cal L}=30\ifb$; we use the double $b$-tagging option.
For $m_H=110,120\GeV$, in Table~\ref{Table5p} we show the rescaled values for
${\cal L}=10\ifb$, along with the improved values using the $I$-factor in
Table~\ref{Table4}, and the required integrated luminosity for discovery
($5\sigma$). We can see that the DiTevatron would see evidence ($3\sigma$)
for Higgs-boson masses up to $m_H=120\GeV$ with ${\cal L}=10\ifb$ (short-term)
and would discover Higgs bosons ($5\sigma$) up to the same mass with
${\cal L}=25\ifb$ (long-term). For comparison, at the Tevatron the significance
of the $m_H=120\GeV$ signal is 1.7 before the $\cos(\theta^*)$ cut and 2.1
after the cut. Therefore, $57\ifb$ would be required to achieve a $5\sigma$
significance.

\begin{table}[t]
\caption{Number of signal ($N_{\rm S}$) and background ($N_{\rm BG}$) events at
the DiTevatron with ${\cal L}=10\,{\rm fb}^{-1}$ for associated Higgs-boson
production and decay through the $b\bar b$ mode. We also show the statistical
significance (${\rm Sig}=N_{\rm S}/(N_{\rm BG})^{1/2}$), the improvement factor
($I$) from Table~5, the corresponding improved signicance ($\rm Sig^I$), and
the required integrated luminosity for discovery ($5\sigma$).}
\label{Table5p}
\begin{center}
\begin{tabular}{|c|c|c|c|c|c|c|}\hline
$m_H$&$N_{\rm S}$&$N_{\rm BG}$&Sig&$I$&$\rm Sig^I$&${\cal L}(5\sigma)$\\ \hline
$110$&32&110&3.1&1.19&3.6&$19\ifb$\\
$120$&23&87&2.4&1.25&3.1&$26\ifb$\\ \hline
\end{tabular}
\end{center}
\hrule
\end{table}

We should add that in addition to the $H\to b\bar b$ mode, it has been
recently pointed out \cite{Kane} that the $H\to\tau^+\tau^-$ mode could
be used to increase the significance of the Higgs signal. We have also
studied this mode and find the signal to be small once currently available
experimental data are used to determine the expected dectection efficiencies.
Moreover, the expected backgrounds are large and it appears difficult to reduce
them enough to obtain a statistically meaningful result. A summary of this
analysis is given in Appendix~C.

As we have indicated above, the above mass reach ($m_H\lsim120\GeV$) applies to
the lightest Higgs boson ($h$) in the supergravity models also. Moreover, in
this case there is an upper limit on $m_h$ which depends on $m_t$:
$m_h\lsim120\,(130)\GeV$ for $m_t=150\,(170)\GeV$. Therefore, Higgs searches at
the DiTevatron would probe a large fraction (if not all) of the parameter space
of the various models which we have considered. In contrast, the reach of LEPII
for Higgs-boson searches is roughly $\sqrt{s}-95\GeV$ \cite{Sopczak}. With a
beam energy of $\sqrt{s}=190\GeV$, the mass reach would be $m_h\lsim95\GeV$.

\section{Discussion and conclusions}

Let us now contrast the potential of the Tevatron versus the DiTevatron for
probing the parameter space of the models we consider. For trileptons searches,
the $\xi_0=0$ case of the conservative models gives the largest rates, which
decrease for increasing values of $\xi_0$. The reaches for various values of
$\xi_0$ and $\tan\beta$ are summarized in Table~\ref{Table3}. Generally
speaking, unless the Tevatron can already  reach all of the allowed range of
chargino masses, the DiTevatron results in an increase of 5--15 GeV in the
reach for chargino masses, for the same integrated luminosity. The increases
are larger for $\tan\beta=2$ and the smaller values of $\xi_0$.

In the minimal $SU(5)$ supergravity model the rates are smaller (since
$\xi_0\gsim3$), but a much smaller range of chargino masses needs to be
explored (see Eq.~(\ref{minranges})), and both the Tevatron and DiTevatron
would explore {\em all} of the allowed parameter space with ${\cal L}=25\ifb$.
The no-scale $SU(5)\times U(1)$ supergravity model, essentially a $\xi_0=0$
model, predicts somewhat larger trilepton rates than its $\xi_0=0$ conservative
model counterpart, and likely constitutes the upper limit for trilepton rates
in supergravity models. The reach in this case would be (for $\mu<0$)
$m_{\chi^\pm_1}\approx190\,(200)\GeV$ at the Tevatron and $220\,(240)\GeV$ at
the DiTevatron, for ${\cal L}=10\,(25)\ifb$.

We should point out that the sensitivity of the DiTevatron for trileptons
would be enhanced if the tracking and calorimeter coverage are improved
relative to those in the present CDF detector. This effect is reflected in
the analysis by the drop in acceptance from 12\% down to 9\% when going from 2
to 4~TeV.

For gluino and squark searches, the reaches are summarized in
Table~\ref{Table5} for the two extreme scenarios of heavy squarks, and
comparable squark and gluino masses. We conclude that at the Tevatron the reach
for gluinos would not exceed $\sim430\,(440)\GeV$ with  ${\cal
L}=10\,(25)\ifb$. Considering that squarks and gluinos can be as massive as
$\sim1\TeV$, this is a modest ($\sim45\%$) reach into parameter space. At the
DiTevatron the reach would improve significantly: $m_{\tilde g}\approx
m_{\tilde q}\sim720\,(750)\GeV$ with ${\cal L}=10\,(25)\ifb$, \ie, $\sim75\%$
of the parameter space. Equation~(\ref{minranges}) shows that {\em all} of
the parameter space of minimal $SU(5)$ model could be explored for both
machines. Note that the reach for gluinos and squarks at the DiTevatron is
considerably larger than the corresponding reach at the Tevatron. This
significant improvement is due to that the Tevatron is at the phase space
limit for squark and gluino production.

For Higgs searches through the $b\bar b$ mode the reaches are summarized in
Table~\ref{Table5p}. Incorporating our $\cos(\theta^*)$ cut, at the DiTevatron
one could see evidence ($3\sigma$) for Higgs boson  masses as high as 120 GeV
with ${\cal L}=10\ifb$ (short-term) and discover ($5\sigma$) Higgs bosons
up to 120 GeV with ${\cal L}=25\ifb$ (long-term). In the supergravity models
that we consider, these results also apply to the lightest supersymmetric Higgs
boson, whose mass is however, bounded above by $m_h\lsim120-130\GeV$.

In summary, we conclude that the DiTevatron with the luminosity level provided
by the Main Injector is a superior machine compared to the Tevatron for the
same luminosity, {\em as far as the search for supersymmmetry is
concerned}. To make this point apparent in a concise way, we have tabulated
all the relevant physics results in Table~7. In this table we
also show numbers of events for some interesting Standard Model processes
which have a bearing in the search for supersymmetry for practical purposes,
\ie, calibration and precise determination of backgrounds.
\section*{Acknowledgments}
This work has been supported in part by DOE grant DE-FG05-91-ER-40633. We
would like to thank Dick Arnowitt for many suggestions and comments and for
reading the manuscript. We would also like to thank Howie Baer and Xerxes Tata
for useful discussions.

\newpage

\vbox{
\parskip   .7ex
\parindent .8cm
\textwidth 7.2in
\textheight 9.7in
\oddsidemargin -0.37in
\evensidemargin -0.37in
\topmargin -0.7in

\vspace*{-1in}
\noindent Table 7: Detected signal (background) for processes of major physics
interest at the Tevatron and DiTevatron. ${\cal L}=10\ifb=3 \times 10^{32}
\times2\,{\rm years}\, @\, 50\%$.
Note: $r = \sigma({\rm 4 TeV}) /\sigma({\rm 2 TeV}).
$ A asterisk indicates that the CDF detector coverage is assumed.
$$
\begin{array}{lccl}
\hline \hline
\sqrt{s} \; (p \bar{p}) & \mbox{2 TeV} & \mbox{4 TeV} &
			\multicolumn{1}{c}{ \mbox{Comments~~~~} }\\
\hline
 &	& &	\\
\bullet \mbox{{\bf Top quark ($\ast$):}} & & & \\
t \bar{t} \; \mbox{($m_t$ = 170 \mgev)} &
	\mbox{7.6~K~(3.8~K)} & \mbox{46~K~(23~K)}   & r = 6 \\
 &	& & \mbox{based on Ref.~\cite{CDF} }\\
 &	& &	\\
\bullet \mbox{{\bf Higgs boson:}} &	& &	\\
  W+H \rightarrow \ell \nu + b \bar{b} & & 	\\
  \mbox{~~$m_{H}=$ 100 \mgev} &	27~(120) &	44~(142) &
		\mbox{Refs.~\cite{SMW,GH,SMWII} }\\
  \mbox{~~$m_{H}=$ 120 \mgev} &	14~(~71) &	23~(~87) &
		\mbox{Refs.~\cite{SMW,GH,SMWII} }\\
  \mbox{~~$m_{H}=$ 120 \mgev} &	11~(~25) &	17~(~30) &
	\mbox{Refs.~\cite{SMW,GH,SMWII} with $\cos(\theta^{\ast}_{bb})$ cut} \\
 &	& &	\\
\bullet \mbox{{\bf Weak bosons ($\ast$):}} &	& & 	\\
  W\rightarrow\ell \nu & \mbox{13.1~M} & \mbox{28.8~M} &
						r = 2.2 \\
  Z\rightarrow\ell \ell & \mbox{~1.4~M} & \mbox{~3.1~M} &
						r = 2.2 \\
 &	& &	\\
\bullet \mbox{{\bf Dibosons ($\ast$):}} & 	& & \\
  WW\rightarrow2\ell+2\nu & 	156 &		452 & r = 2.9 \\
  WZ\rightarrow3\ell+1\nu &	106 &		297 & r = 2.8 \\
  ZZ\rightarrow4\ell     &	~16 &		~42 & r = 2.6 \\
  W\gamma\rightarrow\ell \nu + \gamma     &
  \mbox{13.6~K~(6.2~K)} & \mbox{29.9~K~(13.6~K)} &
	\mbox{$r = 2.2$; $E_{t}^{\;\gamma} >$ 7 GeV,
	$\Delta R_{\ell \gamma} > 0.7$} \\
  Z\gamma\rightarrow\ell \ell + \gamma     &
  \mbox{~6.4~K~(0.4~K)} & \mbox{12.2~K~(~0.8~K)} &
	\mbox{$r = 1.9$; $E_{t}^{\;\gamma} >$ 7 GeV,
	$\Delta R_{\ell \gamma} > 0.7$} \\
 &	& &	\\
\bullet \mbox{{\bf Supersymmetry:}} &		& &	\\
  W+h \rightarrow \ell \nu + b \bar{b} & & 	\\
  \mbox{~~$m_{h}=$ 120 \mgev} &	11~(~25) &	17~(~30) &
	\mbox{Ref.~\cite{SMW,GH,SMWII} with $\cos(\theta^{\ast}_{bb})$ cut} \\
& & & \mbox{SM-like couplings} \\
& & & \\
  \chichi \; (\rightarrow \ell \ell \ell X) \; (\ast) & & & \\
 \mbox{~~$m_{\chione}$ = 150 \mgev}& 32~(26) & 68~(51) &
		\xi_0 = 1; \; \mu<0,
		\tan{\beta} = 2 \\
 \mbox{~~$m_{\chione}$ = 170 \mgev}& 16~(26) & 39~(51) &
		\xi_0 = 1; \; \mu<0, \tan{\beta} = 2 \\
 \mbox{~~$m_{\chione}$ = 190 \mgev}& 6~(26) & 17~(51) &
		\xi_0 = 1; \; \mu<0, \tan{\beta} = 2 \\
& & & \\
\gluino \gluino, \; \gluino \squark, \; \squark \squark \;
	&  \mbox{$3j+\met$} &   \mbox{$4j+\met$} & \mbox{Jet $P_t>40\GeV$}\\
\;\;\;\;\;\;\;\;\;\; (\rightarrow n\mbox{-jets} + \met) &
				\mbox{$>$ 100 GeV} &  \mbox{$>$ 150 GeV} & \\
  \mbox{(a) $m_{\squark} = m_{\gluino} -$10 \mgev} & & & \\
\mbox{~~$m_{\gluino} = $ 400 \mgev} &
	270~(3700)& 11300~(4350) & N_{\rm BG} = 5 \times N_{Z\rightarrow\nu\nu} \\
\mbox{~~$m_{\gluino} = $ 700 \mgev} &
	~20~(3700)& ~~195~(4350) & N_{\rm BG} = 5 \times N_{Z\rightarrow\nu\nu} \\
 & & & \\
\mbox{(b) $m_{\squark}=$ 1000 \mgev} & & & \\
\mbox{~~$m_{\gluino}=$ 300 \mgev} & 602~(3700) & 5600~(4350) &
		\mbox{$N_{\rm BG} = 5\times N_{Z\rightarrow\nu\nu}$} \\
\mbox{~~$m_{\gluino}=$ 400 \mgev} & ~79~(3700) & 2170~(4350) &
		\mbox{$N_{\rm BG} = 5\times N_{Z\rightarrow\nu\nu}$} \\
\mbox{~~$m_{\gluino}=$ 500 \mgev} & ~~5~(3700) & ~580~(4350) &
		\mbox{$N_{\rm BG} = 5\times N_{Z\rightarrow\nu\nu}$} \\
\hline \hline
\end{array}
$$

}
\newpage

\appendix
\section{The DiTevatron}
	The Tevatron is the highest energy collider in the world today.  The recently
reported evidence for the top quark was only possible because of the energy
reach of the  Tevatron: its discovery at a lower collision energy would have
been unthinkable, even with arbitrarily high luminosity.  The Tevatron's single
magnet ring produces collisions of protons and antiprotons at
$\sqrt{s}=1.8\TeV$.  The superconducting magnets of the ring operate at a field
strength of 4.1 Tesla at the peak beam energy of 900 GeV.  The Tevatron is
itself an upgrade of the original Main Ring at Fermilab, which accelerated
beams of protons to 400 GeV for fixed target experiments.  The luminosity of
the Tevatron is currently being upgraded to $10^{32}\cm^{-2}\s^{-1}$.

	The crowning success of the ill-fated Superconducting Super Collider was the
development to production readiness of a 6.5 Tesla superconducting dipole and
corresponding quadrupole.  A string of these magnets were operated successfully
at this field at 4.2$^\circ$K, validating the magnet technology required for
SSC.  The same magnets were also operated at  2$^\circ$K, producing a field of
8.8 Tesla.  A ring of such magnets, placed in the existing Fermilab tunnel,
could use the same  source and the Tevatron as  injector, and produce
collisions at $\sqrt{s}=4\TeV$ - the DiTevatron.  The beams
adiabatically damp as they are accelerated, so that the DiTevatron luminosity
would be $3\times10^{32}\cm^{-2}\s^{-1}$. The DiTevatron makes it possible to
envision doubling the energy of the Tevatron, with no new tunnel construction,
no magnet R\&D, and no new detectors. It is estimated to cost \$250 million,
and would require $\sim4$ years to build.

	This remarkable opportunity is the result of four happy circumstances.  First,
the Tevatron magnet ring and the DiTevatron ring could be situated in the
existing tunnel compatibly.  Figure~\ref{A1} shows the tunnel cross section in
which the Tevatron has been moved up; the DiTevatron is located on the tunnel
floor, preserving the same beam elevation through the collider experiments; and
the magnet elements required for transfer of beams for fixed-target physics are
routed over the two rings.

	Second, the Tevatron can be used to advantage as a high-energy injector for
the new ring.  Many of the most challenging requirements on the superconducting
magnets concern its field quality at injection energy $E_0$.
Figure~\ref{A2}(a) shows the measured field distribution in the superconducting
magnets of the Tevatron.  Its effective full aperture for colliding beams is 5
cm.  Figure~\ref{A2}(b) shows the measured field distribution in the
superconducting magnets for the SSC.  The sextupole term in $B_y$ was
introduced by design and can be removed straightforwardly.  By the same
criteria of field quality, the effective full aperture of the SSC magnet for
colliding beams is 3 cm.  The beam size damps as $1/\sqrt{E_0}$ , so the higher
the injection energy the less is the required aperture of the magnets.
Thus, with injection to DiTevatron at 400 GeV compared to injection to
Tevatron at 150 GeV, the 5 cm Tevatron aperture maps to a DiTevatron aperture
requirement of $\sqrt{150/400}\times 5\cm=3\cm$. The SSC magnets are thus
adequate for DiTevatron use substantially as-is.

	Also at injection, the persistent currents in the superconducting cables of
the magnet produce error multipole fields which can dilute the beams'
brightness before they can be accelerated.  Figure~\ref{A3} shows the magnitude
of these multipoles for the magnets of the Tevatron.  As indicated, these
multipoles would be negligible at the field strength corresponding to 400 GeV
injection to the DiTevatron.

	Third, the forces on the conductors in the SSC magnets are still under
suitable levels of preload compression at a field strength of 8.8 Tesla.
Figure~\ref{A4} shows the measured stress in the conductor package, as a
function of field strength.  The coils of superconducting magnets are assembled
with a preloaded compressive stress which must be greater than the maximum
Lorentz stress produced at full field; otherwise coil motion and quenching
would occur when the direction of net stress reversed.  Although the SSC
magnets were designed to operate at 6.5 Tesla, its mechanical design contains
sufficient prestress to support operation at 8.8 Tesla.

	Lastly, the cryogenic requirement to operate the DiTevatron ring at
2$^\circ$K would require an additional refrigeration loop in the current
Fermilab cryogenic plant (which operates at 4$^\circ$K), but would not pose a
major additional overall refrigeration load.  The present Tevatron magnets have
a heat load which is $\sim10$ times greater than that of the SSC magnets.  The
additional load of the DiTevatron, cooling at 2$^\circ$K, would present a
$\sim20\%$ increase in the aggregate cooling power.

\newpage

\section{$t \bar{t}$ background in $\tilde g,\tilde q$ searches}
We have studied the missing $E_t$ ($\met$) signal with multi-jets from
$t \bar{t} \rightarrow \ell \; \nu + \mbox{$n$-jets}$
and $Z \rightarrow \nu \nu$
for the signal events ($\gluino \gluino$, $\gluino \squark$, $\squark
\squark$). We used ISAJET V7.06 to generate the signal events; Table~8
summarizes the cross sections obtained by ISAJET.
$$
\begin{array}{lcl}
\multicolumn{3}{c}{\mbox{Table 8: Cross sections at $\sqrt{s}$ = 4 TeV} }\\
\hline \hline
\mbox{Physics Process} & \mbox{$\sigma$ [pb]} &
	\multicolumn{1}{c}{ \mbox{Comments} }\\
\hline
\mbox{$B1$: $Z \rightarrow \nu \nu$} & 152  & \mbox{$P_t(Z) \geq$ 40 GeV} \\
\mbox{$B2$: $t \bar{t} \rightarrow \ell \nu$ + $n$-jets} & 12 &
		\mbox{$m_t$ = 170 \mgev; $\ell = e, \mu\;\;$
				(cf.~$\sigma_{t \bar{t}}^{tot}$ = 42~pb)} \\
\mbox{$S1$: $\gluino \gluino$}  & 0.54 &
			\tan \beta = 4, \mu = -400, m_A = 500, A_t = -100 \\
\mbox{$S2$: $\gluino \gluino + \gluino \squark + \squark \squark$} & 5.8 &
			\tan \beta = 4, \mu = -400, m_A = 500, A_t = -100 \\
\hline \hline
\end{array}
$$
\par\noindent {\bf Note:}
$S1$:~$m_{\gluino}$ = 400 \mgev\sp and $m_{\squark}$ = 800 \mgev;
$S2$:~$m_{\gluino}$ = 400 \mgev\sp and $m_{\squark}$ = 390 \mgev.
$m_{\tilde{b}}$ = $m_{\tilde{t}}$ = $m_{\squark}$ for both cases.
\\[-.1in]

We use a CDF detector simulation package (QFL) and a set of the CDF off-line
codes for the lepton/jet finding and $\met$ calculation. Thus, the $\met$
calculation includes the effect of uninstrumented regions of the detector as
well as the detector smearing. Figure~\ref{Et}(a) shows the $\met$
distributions for $Z \rightarrow \nu \nu$ (solid line), $t \bar{t}$ (dashed
line), $\gluino \gluino$ (dotted line) with $m_{\gluino}$ = 400 \mgev\sp and
$m_{\squark}$ = 800 \mgev, and $\gluino \gluino + \gluino \squark + \squark
\squark$ (dash-dotted line) with $m_{\gluino}$ = 400 \mgev\sp and $m_{\squark}$
= 390 \mgev. Since $\met$ values above $\sim$70 GeV are reliable without
detailed correction from experience in the CDF experiment, we simply set the
$\met$ cut at 100 GeV.

In order to reduce the background, we choose an optimized selection
criteria for jets: $(i)$~$N_{\rm jet} \geq 4$ and $(ii)$~$\Sigma E_t({\rm jet})
\geq$ 300 GeV, where $E_t^{\; min}({\rm jet})$ = 20 GeV. We note that the
identified leptons are not used in the above jet selection, nor are they used
to veto the event. This should give us a conservative estimate on the $t
\bar{t}$ background size. Figure~\ref{Et}(b) shows the $\met$ distributions for
signals and backgrounds after the event selection. The cross section ($\met
\geq$ 100 GeV) and its significance for each physics process are listed in
Table~9.\\[-.1in]
$$
\begin{array}{lccc}
\multicolumn{4}{c}{\mbox{Table 9: Significance} }\\
\hline \hline
\mbox{Physics Process} & \mbox{$\sigma$ [fb]} & N_{event} &
	N_{\rm S}/\sqrt{N_{\rm BG}} \\
	&	\mbox{(after cuts)} & \mbox{@10 fb$^{-1}$} &
	({\rm BG} = B1+B2) \\
\hline
\mbox{$B1$: $Z \rightarrow \nu \nu$} & ~97 & ~970 & n/a \\
\mbox{$B2$: $t \bar{t} \rightarrow \ell \nu$ + $n$-jets} & 142 & 1420 & n/a \\
\mbox{$S1$: $\gluino \gluino$}  & 220 & 2200 & ~45 \\
\mbox{$S2$: $\gluino \gluino + \gluino \squark + \squark \squark$}
				& 922 & 9220 & 189 \\
\hline \hline
\end{array}
$$
As one can see, the $t \bar{t}$ background is comparable to the
$Z \rightarrow \nu \nu$ background, even without the lepton removal.
Therefore, our conservative background estimate in the text, \ie,
$5 \times N_{Z \rightarrow \nu\nu}$, is fairly safe.

If we require to remove the events where the leptons are lost, \ie, isolated
($Iso < 4$ GeV) leptons in a CDF fiducial detector region with $P_{t}(\ell)
\geq 15$ GeV ($|\eta(\ell)| < 1.2$), then the $t \bar{t}$ background reduces by
24\% -- not a major improvement. Here the isolation variable ($Iso$) is defined
to be the sum of the transverse energy (excluding the lepton $E_t$) within a
cone of $R = 0.4$ around the lepton.  The two main reasons for the smaller than
expected reduction in the $t\bar t$ background are:
\begin{description}
\item (a) The lepton from $W$ decay ($t\to Wb\to l\nu_l b$) is relatively soft
after the jet activity selection and $\met$ cut;
\item (b) The pseudorapidity region for $e$ and $\mu$ is not wide because we
assumed the specifics of the present CDF detector (\ie, the CDF central
tracking volume).
\end{description}

In summary, $t \bar{t}$ events are not really the major background
for the squark and gluino signals, even if we take the worse case scenario
where the present CDF detector is used without any improvements on the
tracking, and the muon and electron detection coverage.
\newpage

\section{The $H\to\tau^+\tau^-$ signal at the DiTevatron}
We have studied the Standard Model Higgs signals in
$p \bar{p} \rightarrow W+H \; (Z+H) \rightarrow jj + \tau \tau$ at
$\sqrt{s}$ = 4 TeV. We are especially interested in $m_H=120\GeV$,
because the discovery sensitivity at 120 \mgev\sp is the minimum detectable
one and it could be enhanced by adding the $\tau\tau$ mode. In what follows
we fix $m_H=120\GeV$. The cross sections for the associated production of
Higgs, $W+H$ and $Z+H$, are $\sigma_{WH}$ = 440~fb and $\sigma_{ZH}$ = 230~fb
\cite{GH}. The branching ratios are:
\begin{itemize}
\item $B(H \rightarrow \tau \tau)$ = 7\%
\item $B(\tau \rightarrow \mbox{hadrons})$ = 63.9\%, $\;$
	$B(\tau \rightarrow \ell)$ = 36.1\% ($\ell$ = $e, \; \mu$)
\item $B(W \rightarrow jj)$ = 68.5\%, $\;$ $B(Z \rightarrow jj)$ = 69.8\%
\end{itemize}
We apply the following kinematical and geometical cuts:
\begin{itemize}
	\item $P_t(\tau) \geq$ 20 GeV, $|\eta(\tau)| < 2$
	\item $P_t(j) \geq$ 20 GeV, $|\eta(j)| < 2$
	\item $\Delta R(\tau \tau) > 0.7$
	\item $\Delta R(\tau j) > 0.7$
\end{itemize}
Note that $\sigma/E(j)$ = $80\%/\sqrt{E} \oplus 5\%$
and $\sigma/E(\tau)$ = $30\%/\sqrt{E} \oplus 3\%$.
The geometical and kinematical acceptance (${\cal A}$) is obtained to be 19\%
using PYTHIA \cite{pythia}.

As for $\tau$ identification ($P_{t}(\tau) \geq$ 20 GeV),
we simply refer to the selection
in the CDF data analyses:\\

\par\noindent \underline{$\tau \rightarrow \mbox{hadrons}$}~\cite{T8}
\begin{itemize}
	\item $\tau$ reconstruction efficiency: 94\%
		\begin{itemize}
		\item Seed track $P_{t} \geq 5$ GeV
		\item Clustering based on tracks ($P_{t} \geq 1$ GeV)
			within a $30^{\circ}$ cone around the seed track
		\item $E_{t} \geq 15$ GeV with
			$E_{em}/(E_{em}+E_{had}) < 0.95$
		\end{itemize}
	\item $\Sigma{P_{t}}$ cut efficiency: 86\%
		\begin{itemize}
		\item $\Sigma{P_{t}}$ = $\Sigma{P_{t}(tracks)} +$
			$\Sigma{ E_{t}( \mbox{$\pi^{0}$'s} ) }$
		\item $\Sigma{P_{t}} \geq$ 17.5, 20 or 22.5 GeV for
			1, 2 or 3 prong.
		\end{itemize}
	\item Isolation cut efficiency: 84\%
		\begin{itemize}
		\item No tracks between 10 and $30^{\circ}$ from
			the seed track.
		\item This efficiency is estimated from
			$W\to e\nu_e$ data.
			The loss is because underlying event
			tracks overlap with the electron.
		\end{itemize}
	\item $N(tracks)$ cut efficiency: 98\%
		\begin{itemize}
		\item Number of tracks should be $\leq 3$ in the $10^{\circ}$
			cone.
		\end{itemize}
\end{itemize}
The total efficiency is 67\% per $\tau$. In this analysis we use
\begin{equation}
	\epsilon^{ID}_{\tau \rightarrow h}  =  70\% \nonumber
\end{equation}
The probability for a QCD jet to satisfy the $\tau$ selection is estimated
to be 0.7\% \cite{T8}.\\

\par\noindent \underline{$\tau \rightarrow \ell \nu \nu$}
\begin{itemize}
	\item $\ell$ (from $\tau$) identification efficiency: 62\%
		\begin{itemize}
		\item Kinematical acceptance is 69\% \cite{pythia}
		with $E_{t}(e) \geq 10$ GeV or $P_{t}(\mu) \geq 10$ GeV
		for $P_{t}(\tau) \geq 20$ GeV.
		\item Electron and muon quality cut efficiency is 90\%.
		\end{itemize}
	\item Isolation cut efficiency: 84\% \cite{T8}
		\begin{itemize}
		\item No tracks between 10 and $30^{\circ}$ from
			the seed track.
		\item This efficiency is estimated from
			$W \rightarrow e \; \nu$ data.
			The loss is because underlying event
			tracks overlap with the electron.
		\end{itemize}
	\item $N(tracks)$ cut efficiency: 98\% \cite{T8}
		\begin{itemize}
		\item Number of tracks should be 1 in the $10^{\circ}$
			cone.
		\end{itemize}
\end{itemize}
The total efficiency is 51\% per $\tau$. In this analysis we use
\begin{equation}
	\epsilon^{ID}_{\tau \rightarrow \ell}  = 50\% \nonumber
\end{equation}

The number of events is calculated as follows:
\begin{eqnarray}
 N_{VH} & = & \sigma_{VH} \times {\cal L} \times B(V \rightarrow jj)
	\times B(H \rightarrow \tau \tau)
	\times {\cal A} \nonumber \\
	& & \times B(\tau \rightarrow x)
	\times B(\tau \rightarrow y) \times N_{xy}(combination)
	\times \epsilon_{x}^{ID}
	\times \epsilon_{y}^{ID}
\end{eqnarray}
where $V$ is $W$ or $Z$;
${\cal L}$ is the integrated luminosity;
${\cal A}$ is the geometical and kinematical acceptance;
$x$ ($y$) refers to $\tau$ leptonic or hadronic decay mode;
$N_{xy}$ is the number of combinations for a choice of $x$ and $y$ decay
modes; $\epsilon^{ID}$ is the $\tau$ identification efficiency.
Table 10 summarizes the number of events expected at an integrated luminosity
of 10~fb$^{-1}$. It should be noted that these numbers are obtained for a fully
instrumented detector in $| \eta | < 2$ and 100\% trigger efficiency, \ie,
they are slightly optimistic.

$$
\begin{array}{lcccc}
\multicolumn{5}{c}{\mbox{Table 10: Number of events for $m_H=120\GeV$ with
10~fb$^{-1}$} } \\
\hline \hline
\multicolumn{1}{c}{\mbox{Mode ($x$, $y$)}} & N_{xy} &
					W+H & Z+H & \mbox{Total} \\
\hline
\tau \rightarrow \mbox{hadrons}, \tau \rightarrow \mbox{hadrons} & 1 &
		8.0	& 4.3	& 12 \\
\tau \rightarrow \mbox{hadrons}, \tau \rightarrow \ell & 2 &
		6.5 	& 3.4 	& 10 \\
\tau \rightarrow \ell, \tau \rightarrow \ell & 1 &
		1.3	& 0.7 	& ~2 \\
\hline
\hline
\end{array}
$$

At this level, the expected sizable Standard Model backgrounds are:
QCD 4 jets ($\rightarrow$ ``$\tau \tau$'' + $j \; j$),
$Z (\rightarrow \tau \tau)$ + 2-jets,
$WZ \rightarrow jj + \tau \tau$,
$ZZ \rightarrow jj + \tau \tau$ ($\tau \tau + jj$),
$t \bar{t} \rightarrow \tau^{+} \nu b + \tau^{-} \bar{\nu} \bar{b}$,
Drell-Yan $\tau \tau$ + 2-jets, $etc$.

In the decay of $H \rightarrow \tau \tau$, the azimuthal opening angle of two
$\tau$'s is often near 180$^{\circ}$. Therefore, the missing $E_t$ ($\met$) is
soft. A further cut on $M_{jj}$ ($60 < M_{jj} < 110$ \mgev) can
reduce the QCD and $t \bar{t} \rightarrow \tau^{+} \nu b + \tau^{-} \bar{\nu}
\bar{b}$ (460 fb just in cross section times branching ratio) backgrounds
while keeping most of the signals in Table~10.

Now let us consider the remaining backgrounds for each signal mode and give
a simple estimate of their sizes and possible cuts to reduce them:
\begin{itemize}
\item $\underline{\tau \rightarrow \mbox{hadrons},
		\tau \rightarrow \mbox{hadrons}}$\\
	This mode will be the best to determine the mass of the Higgs boson.
	However, it suffers from QCD jets background.
	The QCD dijet cross section is 27 $\mu$b
	for $30 < P_t < 70$ GeV \cite{Isajet},
	where the dijet invariant mass is
	near the weak boson masses.
	By requiring two more jets ($E_{t} > 20$ GeV),
	the cross section is about 1\% of 27~$\mu$b
	(${\cal O}(\alpha^2_s)$), \ie, 270~nb.
	The fake rate is determined to be 0.7\% by CDF \cite{T8}.
	By taking into account 6 combinations (2 out of 4 jets)
	in misidentification of 2 jets as 2 $\tau$'s,
	the cross section for ``$\tau \tau$'' + $jj$-like events
	is $79 \times 10^{3}$~fb.
	With 10~fb$^{-1}$,
	we have to reduce the background (790~K events) to 16 (6) events
	for a $3 \sigma$ ($5 \sigma$) significance.
	It should be noted that the $\met$ in the events is not hard.
	Therefore, the most efficient requirement is
	a mass cut, $M_{\tau \tau} \geq 100$ GeV,
	which will keep a large fraction of the signal.
	Though the cut reduces $Z$, $WZ$ and $ZZ$ backgrounds substantially,
	the QCD jet events as well as $t \bar{t}$ events
	will remain as the major backgrounds
	because the mass distributions in those backgrounds
	are continuum and broad.
	If we want to reduce the QCD background, the $\met$ cut should be
	applied ($\met \geq 20$ GeV). This will kill the signal.
	If we want to reduce the $t \bar{t}$ background, the $\met$ cut should
	be $\met \leq 20$ GeV, and we suffer then from the QCD background.
	Thus, we find that it would be very hard to see a statistically
	significant Higgs signal at 10~fb$^{-1}$.
\item $\underline{\tau \rightarrow \mbox{hadrons}, \tau \rightarrow \ell}$ \\
	This event topology ($\ell$ + $\tau$-jet+ 2 jets) is expected from
	4 jets (at least one jet contains heavy flavours, \eg,
	$g \rightarrow b \bar{b}$),
	$Z$ + 2 jets, $WZ/ZZ$, and $t\bar t$ events.
	To reduce the QCD 4-jet background we need to require
	a higher $P_t(\ell)$ cut (\eg, 20 GeV) and $\met$ cut at 20 GeV.
	The higher lepton $P_t$ cut will reduce the signal by $\sim$35\%
	in 120-GeV Higgs-boson decay, that is, down to 6-7 events.
	The $\met$ cut reduces the signal further.
	We also need a cut on the transverse mass distribution of the
	$\ell + \mbox{$\tau$-jet} + \met$ system ($e.g.$ $\geq$ 90 GeV)
	to remove $Z$ + 2 jets and $WZ/ZZ$ events. As for $t \bar{t}$ events,
	cuts on $P_{t}(\ell)$, $\met$, and the transverse mass will not be
	efficient to reduce this background.
	Since the mass spectrum is very wide,
	the determination of the Higgs-boson mass is difficult with
	10~fb$^{-1}$
	even if we can achieve {\bf zero} background without
	loosing any signal events from the transverse mass cut.
\item $\underline{\tau \rightarrow \ell, \tau \rightarrow \ell}$ \\
	This mode is expected to be cleaner (2 isolated leptons + 2 jets).
	However, this event topology is also expected
        from Drell-Yan ($\rightarrow \ell^{+} \ell^{-}$) + 2 jets,
	$Z (\rightarrow \ell^{+} \ell^{-})$ + 2 jets, bb + 2 jets,
        WZ/ZZ events, and $t \bar{t}$ events.
	To reduce the backgrounds we need to select $e \mu$ events
	with higher $P_t(\ell)$ cut (\eg, 20 GeV).
	The higher $P_{t}$ cut for lepton is estimated to accept 42\%
	of signals in Table 10,
	so that the signal is reduced by a factor of $\sim$4
	by excluding the $ee$ and $\mu \mu$ modes.
	Therefore, we will see no signal with 10~fb$^{-1}$.
\end{itemize}
In conclusion, we expect it to be very difficult to get a significant signal
above the background in the $\tau\tau$ mode with 10~fb$^{-1}$.

Since a track isolation cut is essential for $\tau$ identification,
we should operate an accelerator machine with as few interactions per
beam crossing as possible. The current CDF data indicates
84\% in its efficiency, and the loss is because underlying event
tracks overlap with $\tau$ tracks \cite{T8}.
Therefore, in any high luminosity operation such as at the LHC
(6 interactions per crossing even at 10$^{33}$~cm$^{-2}$s$^{-1}$)
or T$^{\ast}$, this efficiency is expected to be lower.

\newpage

\begin{figure}[p]
\vspace{6.0in}
\includegraphics{DiTev.SSM2.ps}
\caption{Trilepton yield ($\sigma\times B$) versus chargino mass in chargino
production in $p\bar p$ collisions. The lines define the range of parameters
allowed within the most conservative supergravity model with universal
soft-supersymmetry-breaking and radiative electroweak symmetry breaking.
Results are shown for $\xi_0=m_0/m_{1/2}=0,1,2,5$ ($\leftrightarrow m_{\tilde
q}\sim(0.8-2)m_{\tilde g}$); $A=0$, $\tan\beta=2$, and
both signs of the Higgs mixing parameter $\mu$ (we use $m_t^{\rm
pole}=174\,{\rm GeV}$). The upper (lower) plots show the limits which could be
reached at the Tevatron (DiTevatron). The estimated sensitivity limit at the
Tevatron (DiTevatron) for ${\cal L}=10\,{\rm fb}^{-1}$ is 21\,(40)~fb, and for
${\cal L}=25\,{\rm fb}^{-1}$ is 14\,(25)~fb.}
\label{DiTev.SSM2}
\end{figure}
\clearpage

\begin{figure}[p]
\vspace{6.0in}
\includegraphics{DiTev.SSM10.ps}
\caption{Trilepton yield ($\sigma\times B$) versus chargino mass in chargino
production in $p\bar p$ collisions. The lines define the range of parameters
allowed within the most conservative supergravity model with universal
soft-supersymmetry-breaking and radiative electroweak symmetry breaking.
Results are shown for $\xi_0=m_0/m_{1/2}=0,1,2$; $A=0$, $\tan\beta=10$, and
both signs of the Higgs mixing parameter $\mu$ (we use $m_t^{\rm
pole}=174\,{\rm GeV}$). (The corresponding curves for $\xi_0=5$ are not shown
since they largely overlap with those for $\xi_0=2$.) The upper (lower) plots
show the limits which could be reached at the Tevatron (DiTevatron). The
estimated sensitivity limit at the Tevatron (DiTevatron) for ${\cal L}=10\,{\rm
fb}^{-1}$ is 21\,(40)~fb, and for ${\cal L}=25\,{\rm fb}^{-1}$ is 14\,(25)~fb.}
\label{DiTev.SSM10}
\end{figure}
\clearpage

\begin{figure}[p]
\vspace{6.0in}
\includegraphics{DiTev.min.ps}
\caption{Trilepton yield ($\sigma\times B$) versus chargino mass in chargino
production in $p\bar p$ collisions. The dots define the range of parameters
allowed within the minimal $SU(5)$ supergravity model (for $m^{\rm
pole}_t=168\GeV$ and $\tan\beta=2-10$).
Results are shown for each sign of the Higgs mixing parameter $\mu$. The upper
(lower) plots show the limits which could be reached at the Tevatron
(DiTevatron). The estimated sensitivity limit at the Tevatron (DiTevatron) for
${\cal L}=10\,{\rm fb}^{-1}$ is 21\,(40)~fb, and for
${\cal L}=25\,{\rm fb}^{-1}$ is 14\,(25)~fb.}
\label{DiTev.min}
\end{figure}
\clearpage

\begin{figure}[p]
\vspace{6.0in}
\includegraphics{DiTev.nsc.ps}
\caption{Trilepton yield ($\sigma\times B$) versus chargino mass in chargino
production in $p\bar p$ collisions. The dots define the range of parameters
allowed within the string-inspired no-scale $SU(5)\times U(1)$ supergravity
model (for $m^{\rm pole}_t=178\GeV$). Results are shown for each sign of the
Higgs mixing parameter $\mu$. The upper (lower) plots show the limits which
could be reached at the Tevatron (DiTevatron). The estimated sensitivity limit
at the Tevatron (DiTevatron) for ${\cal L}=10\,{\rm fb}^{-1}$ is 21\,(40)~fb,
and for ${\cal L}=25\,{\rm fb}^{-1}$ is 14\,(25)~fb.}
\label{DiTev.nsc}
\end{figure}
\clearpage

\begin{figure}[p]
\vspace{5.0in}
\includegraphics{MonoTev.gluino.ps}
\vspace{1.7in}
\caption{Statistical significance for gluino and squark events at the
{\bf Tevatron} with ${\cal L}=10\,{\rm fb}^{-1}$.  (The significance scales
with ${\cal L}^{1/2}$.) These events were selected by the criteria
$P_t >100\,{\rm GeV}$, and 3 jets with $P_t > 40\,{\rm GeV}$. Bands are
shown for signal S from gluino pairs ($m_{\tilde q}=1\,{\rm TeV}$), and
squark/gluino combinations ($m_{\tilde q}=m_{\tilde g}-10\,{\rm GeV}$). These
two cases bracket the range of possibilities in the conservative supergravity
models which we consider. The background $BG$ is calculated from
$Z\to\nu\bar\nu$; the bands provide for a factor of 5 deterioration of
$N_{\rm S}/N_{\rm BG}$
ratio due to additional backgrounds or inefficiencies.}
\label{MonoTev.gluino}
\end{figure}
\clearpage

\begin{figure}[p]
\vspace{5.0in}
\includegraphics{DiTev.gluino.ps}
\vspace{1.7in}
\caption{Statistical significance for gluino and squark events at the {\bf
DiTevatron} with ${\cal L}=10\,{\rm fb}^{-1}$. (The significance scales with
${\cal L}^{1/2}$.) These events were selected by the criteria
$P_t > 150\,{\rm GeV}$, and 4 jets with $P_t > 40\,{\rm GeV}$.
Bands are shown for signal S from gluino pairs ($m_{\tilde q}=1\,{\rm TeV}$),
and squark/gluino combinations ($m_{\tilde q}=m_{\tilde g}-10\,{\rm GeV}$).
These two cases bracket the range of possibilities in the conservative
supergravity models which we consider. The background $BG$ is calculated from
$Z\to\nu\bar\nu$; the bands provide for a factor of 5 deterioration of
$N_{\rm S}/N_{\rm BG}$
ratio due to additional backgrounds or inefficiencies.}
\label{DiTev.gluino}
\end{figure}
\clearpage

\begin{figure}[p]
\vspace{6.0in}
\includegraphics{costheta.ps}
\vspace{1in}
\caption{The $\cos(\theta^{\ast}_{bb})$ distributions (before any event
selection) for $Z+H(\rightarrow bb)$ ($m_{H} = 120\GeV$) and QCD $Z+bb$ ($30 <
M_{bb} < 200\GeV$) events. The variable $\theta^{\ast}_{bb}$ is the opening
angle between the two $b$-quarks in the center of mass system, where the $z$
axis is defined as the direction of $\vec{p}_{b_1} + \vec{p}_{b_2}$. Note that
the vertical scale is shown as the fraction of events per 0.01.}
\label{costheta}
\end{figure}
\clearpage

\begin{figure}[p]
\vspace{6.0in}
\includegraphics{tunnel.eps}
\caption{ Cross section of Tevatron tunnel, showing arrangement of SSC
magnet (inverted from SSC design), Tevatron magnet, and transfer line magnet.}
\label{A1}
\end{figure}

\begin{figure}[p]
\vspace{6.0in}
\includegraphics{A2.eps}
\caption{Field distribution in Collider magnets and relation to effective
aperture: (a) rms deviations in full production run of Tevatron magnets;
(b) rms deviations in 8 Fermilab-built SSC magnets.  $B_y$ contains a sextupole
term which was built in by design, which would be removed for DiTevatron
magnets.}
\label{A2}
\end{figure}

\begin{figure}[p]
\vspace{6.0in}
\includegraphics{A3.eps}
\caption{Persistent-current sextupole field for a Tevatron magnet, as a
function of current.  400 GeV injection field is indicated.}
\label{A3}
\end{figure}

\begin{figure}[p]
\vspace{6.0in}
\includegraphics{A4.eps}
\caption{Net mechanical stress (sum of preload and Lorentz stress) as a
function of current$^2$.  Peak field of 8.8 Tesla is indicated.}
\label{A4}
\end{figure}

\begin{figure}[p]
\vspace{6.0in}
\includegraphics{Et.ps}
\vspace{1in}
\caption{The $\met$ distributions before (a) and after (b) event selection
for $Z \rightarrow \nu \nu$ (solid line), $t \bar{t}$ (dashed line),
$\gluino \gluino$ (dotted line)
with $m_{\gluino}$ = 400 \mgev\sp and $m_{\squark}$ = 800 \mgev,
$\gluino \gluino + \gluino \squark + \squark \squark$ (dash-dotted line)
with $m_{\gluino}$ = 400 \mgev\sp and $m_{\squark}$ = 390 \mgev. We set the
$\met$ cut at 100 GeV.}
\label{Et}
\end{figure}
\clearpage

\end{document}